\documentclass{article}
\usepackage{spconf,amsmath,graphicx,amssymb, times, multirow}
\ninept
\title{Jointly Tracking and Separating Speech Sources Using Multiple Features and the generalized labeled multi-Bernoulli Framework}
\name{Shoufeng Lin}
\address{Department
of Electrical and Computer Engineering, Curtin University \\ Kent Street, Bentley, Perth, Western Australia, 6102 \\
shoufeng.lin@postgrad.curtin.edu.au }
\begin{document}
%\ninept
%
\maketitle
\begin{abstract}
This paper proposes a novel joint multi-speaker tracking-and-separation method based on the generalized labeled multi-Bernoulli (GLMB) multi-target tracking filter, using sound mixtures recorded by microphones. 
Standard multi-speaker tracking algorithms usually only track speaker locations, and ambiguity occurs when speakers are spatially close. 
The proposed multi-feature GLMB tracking filter treats the set of vectors of associated speaker features (location, pitch and sound) as the multi-target multi-feature observation, characterizes transitioning features with corresponding transition models and overall likelihood function, thus jointly tracks and separates each multi-feature speaker, and addresses the spatial ambiguity problem. 
Numerical evaluation verifies that the proposed method can correctly track locations of multiple speakers and meanwhile separate speech signals. 
\end{abstract}
\begin{keywords}
multi-speaker tracking, multi-feature extraction, speech separation, microphone array processing, GLMB filter.
\end{keywords}

\section{Introduction}
\label{sec:intro}

Multi-speaker tracking using microphones is an important task in smart environments such as automatic camera steering in video conferencing. 
Numerous acoustic multi-speaker tracking algorithms can be found in the literature \cite{ward2003particle, vo2004tracking, ma2006tracking,talantzis2010acoustic}, using various techniques such as mutual information or cross-correlation for spatial localization, and particle filtering for speaker tracking. 
Generic multi-target tracking filters \cite{cphd_vo, vo2013labeled,vo2014labeled, reuter2014labeled} can also be implemented to track multiple speakers online when provided with speaker location estimates as multi-target observations. 
These existing implementations of multi-speaker tracking methods however, usually track only spatial locations of respective speakers. 
Moreover, spatial tracking has the ambiguity problem when speakers are spatially close to each other, because by relying on the location information alone, the tracking filters would take them as a single speaker, hence unable to correctly identify and separate the sound sources in the mixture.

Separating original source signals from the mixtures recorded by microphones has also a wide range of applications such as automatic meeting transcription and speaker recognition. 
Many blind source separation (BSS) methods have been developed \cite{yilmaz2004blind,sawada2004robust, kim2007blind,reju2010underdetermined}, based on the independent component analysis (ICA) or time-frequency masking (TFM) techniques. %, assuming limited or no knowledge of source locations. 
However, it can be challenging for some BSS methods to continuously separate moving sources. Thus the location-based source separation methods, e.g. the wideband beamforming methods \cite{doclo2003design,liu2010wideband}, are often employed as an additional source separation step after obtaining the location tracking results. 
%
%This way of tracking and separating speech sources can usually provide superior speech separation performance compared to the blind speech sepeartion (BSS) methods. 

%In this paper, we propose a novel method to jointly track-and-separate speech sources. 

% Although blind speech separation (BSS) methods do not require knowledge of speaker locations, they often face the challenge in solving the permutation problem and can be sensitive to reverberation \cite{yilmaz2004blind,reju2010underdetermined, sawada2011underdetermined}. 

%
\begin{figure}[!t]
\centering
\includegraphics[width=0.48\textwidth]{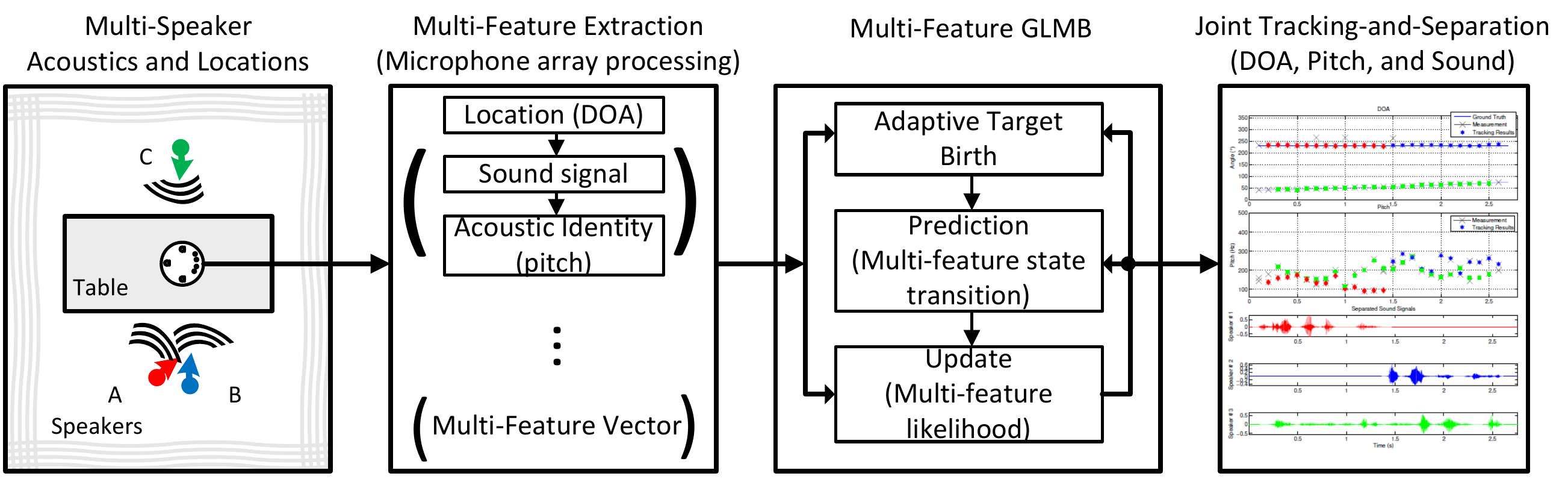}
\centering
\caption{ System overview.}
\label{fig:GLMBmulti_Overview}
\end{figure}

In this paper, we propose a systematic multi-feature tracking-and-separation framework based on the generalized labeled multi-Bernoulli (GLMB) filter \cite{vo2013labeled,vo2014labeled, reuter2014labeled}. 
As shown in Fig.~\ref{fig:GLMBmulti_Overview}, we first obtain multiple speaker features from sound mixtures by detecting locations of all candidate speakers, extracting their corresponding speech signals and estimating the related acoustic identities (pitches).  
Each extracted vector of associated speaker features of a candidate speaker, i.e. the location, pitch and the corresponding speech signals, can be treated as an integral multi-feature target observation. The set of multi-feature vectors forms the multi-target multi-feature observations, which are then tracked in the proposed multi-feature GLMB.
%
%
%to use not only the location but also the acoustic identity (pitch) information of each speaker in the multi-target tracking process, thus when the speakers are closely located, their pitch information can be used to separate them apart, and vice versa. 
%%
%As shown in Fig.~\ref{fig:GLMBmulti_Overview}, we first obtain multiple speaker features from sound signals by detecting locations of all candidate speakers, extracting their corresponding speech signals and estimating the pitches.  
%Each extracted vector of associated speaker features, i.e. the location, the pitch and the corresponding speech signals of each candidate speaker, can be treated as an integral multi-feature target observation, and the set of multi-feature vectors forms the multi-target multi-feature observations, which are then tracked based on the GLMB Bayes framework.
%which is a Bayes random finite set (RFS) multi-target tracker, and consists of a prediction and an update step per each recursion. 
%The proposed method can be viewed as an extension of the GLMB by treating the set of speaker location, identity and sound as the target state, thereby associating the identities of speakers with their locations and sounds. 
Moreover, since the standard implementations of the GLMB framework \cite{vo2013labeled,vo2014labeled, reuter2014labeled} track only one feature, necessary adaptations are required to support multi-feature tracking.
We categorize the location and pitch as ``transitioning'' features, while the non-stationary sound signal as a ``non-transitioning'' feature. 
In the multi-feature GLMB recursion, transitioning features have their own first-order Markov transition models and are directly used for track confirmation in the update step, while the non-transitioning feature is zeroed in the prediction step and assigned with associated extracted sound in the update step. 
We also propose new state transition function and measurement likelihood function for multiple transitioning features. 
The multi-feature GLMB tracking filter produces labeled tracks for respective speakers, the corresponding pitch estimates, as well as the separated sound signals. Furthermore, it also addresses the ambiguity problem because when speakers locate closely, their pitch information can be used to separate them in the multi-feature GLMB tracking algorithm, and vice versa.

%The rest of the paper is organized as follows. Section \ref{sec:features} presents the methods for extracting multiple features of speakers. 
%The multi-feature GLMB is proposed in Section \ref{sec:eGLMB}.
%Numerical results are provided in Section \ref{sec:GLMBmultiResults}, and closing remarks are given in Section \ref{sec:conclusion}.

%I wish you the best of success.

\section{Speaker Feature Extraction}
\label{sec:features}

%Fig.~\ref{fig:GLMBmulti_Overview} gives an overview of the proposed method.
%
%In this paper, we use only one circular microphone array to extract-and-track three features of multiple speakers, i.e. the spatial location (DOA), acoustic identity (pitch) and the sound signal.  
%
%The spatial locations of candidate speakers are estimated using an extension of the generalized cross-correlation - phase transform (GCC-PHAT) method from the microphone array. 
%Then the corresponding sound signals are extracted using the location estimates and a weighted least square (WLS) beamformer. 
%The pitch estimates from sound signals are used as the acoustic identity of the speaker. %, which is obtained from the sound signals using the PEFAC method \cite{gonzalez2014pefac}. 

\subsection{Speaker Localization}
\label{sec:mcc-phat}
We use a circular microphone array in this paper. Denote the sound signals captured by the microphone array as $x_j(t)$ and locations of microphones as $\vec{m}_j$, where $t \in \mathbb{R},~ j=1,..., M$, integer $M$ is the number of microphones. 
%
%Since the direct-path component is usually stronger than random reflections, and also motivated by the idea of using the redundant information from multiple microphones to suppress the random reflections \cite{benesty2004time}, w
We formulate a multi-channel implementation of the generalized cross-correlation - phase transform (GCC-PHAT) method \cite{knapp1976generalized}, which we refer to as the MCC-PHAT:
\begin{equation} \label{eq:mcc-phat}
\xi^{\mathrm{mcc-phat}} (k,\varsigma) \triangleq  \prod \limits_{(i,j) \in P }   \xi_{ij}^{\mathrm{gcc-phat}} (k,\tau_{ij}(\varsigma))  ,
\end{equation}
where
\begin{equation} \label{eq:ximccphat}
\xi_{ij}^{\mathrm{gcc-phat}} (k,\tau_{ij}(\varsigma)) = 
% \mathrm{FFT}^{-1} ( \Xi_{ij}^{\mathrm{gcc-phat}}(k,f) )
\int _{-\infty} ^{+\infty} \Xi_{ij}^{\mathrm{gcc-phat}}(k,f) \cdot e^{\mathrm{i} 2 \pi f \tau_{ij}(\varsigma)}  df ,
\end{equation}
and
\begin{equation} \label{eq:Xi}
\Xi_{ij}^{\mathrm{gcc-phat}}(k,f) = \frac{X_i(k,f) \cdot X_j^{\star}(k,f)}{|X_i(k,f) \cdot X_j^{\star}(k,f)|} .
\end{equation}
Here $\mathrm{i}=\sqrt{-1}$, $[\cdot]^{\star}$ the complex conjugate operation, $X_i(k,f)$ and $X_j(k,f)$ are respectively the short-time Fourier transforms of microphone signals $x_i(\cdot)$ and $x_j(\cdot)$ at time frame $k$. 
(In practice, sound signals are discretized into $x_i(n),~ n \in \mathbb{Z}$ at a sampling frequency $f_s = 48000$Hz, thus the short-time FFT is used in (\ref{eq:Xi}), and the integration in (\ref{eq:ximccphat}) becomes a summation.)

Time difference $\tau_{ij}$ is a function of speaker direction of arrival (DOA) $\varsigma \in [0, 360^{\circ})$ from a distance of $r = 1$m (far-field)
\begin{equation} \label{eq:tauVSp}
\tau_{ij}(\varsigma) = (\| \vec{\wp}(\varsigma) - \vec{m}_i\| - \| \vec{\wp}(\varsigma) -\vec{m}_j\|) /v ,
\end{equation}
\begin{equation} \label{eq:pVStheta}
\vec{\wp} (\varsigma) = [r\cdot \cos \varsigma,~ r\cdot \sin \varsigma] .
\end{equation}
To avoid spatial alias, the set of microphone pairs $P$ is
\begin{equation}
\label{eq:micPair}
P = \{ (i,j) \; | \| \vec{m}_i - \vec{m}_j \| < v/f_{max} ) ;  \; 1 \leq i < j \leq M \} ,
\end{equation}
where $v = 343$m/s is the velocity of sound, and $f_{max} = 3600$Hz is the maximum signal frequency considered. 

In this paper, we use only one circular microphone array in the azimuth plane. (Cartesian locations of speakers can be obtained using multiple microphone arrays.) %Thus the DOA is used as the speaker location.
The set of estimated DOAs of candidate speakers are denoted as $\hat{\Theta}_k$ at time $k$:
\begin{equation} \label{eq:doaset}
\hat{\Theta}_k = \{ \hat{\varsigma}_{k,i} ~|~ {i} = 1,\dots,{N_k} \} ,
\end{equation}
where $\hat{\varsigma}_{k,i}$ correspond to the local peaks of $\xi^{\mathrm{mcc-phat}} (k,\cdot)$, and integer $N_k \geq 0$ denotes the number of detected speakers (accounting for spurious estimates from reflections, and miss detections due to non-stationary or competing speech signals) at frame $k$. 
$N_k=0$ indicates that no candidate speaker is detected and thus $\hat{\Theta}_k = \emptyset$. 
Assuming in general that the spurious  estimates and miss detections exhibit no temporal consistency from one time frame to the next, while the estimates from true speakers follow a kinematic model, tracking filters \cite{ward2003particle, ma2006tracking, vo2013labeled,vo2014labeled, reuter2014labeled} can be applied to track speaker locations.  Such approach is also applied for tracking multiple features as shown in Section \ref{sec:eGLMB}. 
%
%$\hat{\theta}_{k,i}$ satisfies $\bar{\xi}(k,\hat{\theta}_{k,i}) \geq T_{\bar{\xi}}$, ($T_{\bar{\xi}} > 0$ is a threshold), and
%%
%\begin{equation} \label{eq:doa2}
%\hat{\theta}_{k,i} = \argmax _{\theta_{k,i} \in [\hat{\theta}_{k,i} - \theta_{r}, \hat{\theta}_{k,i} + \theta_{r}]} \bar{\xi}(k,\theta_{k,i}) .
%\end{equation}

\subsection{Sound Extraction}

Speech signals from the DOA estimates $\hat{\varsigma}_{k,i}$ can then be extracted from the sound mixtures recorded by microphones. 
Here we implement the wideband weighted least square (WLS) beamforming method \cite{liu2010wideband} for sound extraction.

The WLS beamformer uses the filter-and-sum structure, and has $J_t=32$ taps in each channel. Its mainlobe steers to the speaker DOA $\hat{\varsigma}_{k,i}$, and the corresponding sidelobe ranges from $\hat{\varsigma}_{k,i} + 15^{\circ}$ to $\hat{\varsigma}_{k,i} - 15^{\circ} $. The frequency range used is [20, 8000]Hz.

The real-valued $(J_t \cdot M) \times 1$ optimal weight vector $\mathbf{w}_{k,i} $ for a DOA $\hat{\varsigma}_{k,i}$ is obtained according to the wideband WLS beamformer \cite{liu2010wideband} and using the microphone locations $\vec{m}_j$, then the extracted sound signal at time frame $k$ can be calculated from:
\begin{equation} \label{eq:soundsep}
\hat{s}_{k,i} \big( n ) = \mathbf{w}_{k,i} ^T  ~ \mathbf{x}(n) ,  
\end{equation}
where $[\cdot]^T$ is the matrix transpose, and % $n$ is the index in the frame $k$, 
\begin{equation}
\begin{aligned}
{\mathbf{x}} (n) \! =  \!
\begin{bmatrix}
\mathbf{x}_0(n), \dots,
\mathbf{x}_{j_t}(n) , \dots,
\mathbf{x}_{J_t-1}(n) 
\end{bmatrix} ^T \!\!\!  ,~ j_t \in \! [0, J_t\! - \! 1]
\end{aligned} 
\end{equation}
\begin{equation}
\begin{aligned}
\mathbf{x}_{j_t}( n ) =
\begin{bmatrix}
{x}_1(n + j_t), \dots,
{x}_j(n + j_t), \dots, 
{x}_M(n + j_t)
\end{bmatrix} .  
\end{aligned} 
\end{equation}
%
%$j_t \in [0, J_t -1]$, and $[\cdot]^T$ is the matrix transpose.

\subsection{Acoustic Identity}

The extracted sound $\hat{s}_{k,i}$ that corresponds to a speaker location $\hat{\varsigma}_{k,i}$ can further be used to extract speaker's acoustic identity, e.g. pitch, Gaussian Mixture Model (GMM) \cite{reynolds2002overview} parameters, etc. In this paper we use the pitch as a simple acoustic identity, as pitch can be estimated from a short segment of voiced sound, different speakers usually have different pitch, and pitch of a speaker is usually distributed within a limited range. 
Numerous pitch estimation methods can be found in the literature. % \cite{talkin1995robust, sun2002pitch,lee2012noise, gonzalez2014pefac,wang2017robust}. 
Here we employ the PEFAC (Pitch Estimation Filter with Amplitude Compression) method \cite{gonzalez2014pefac} and use the averaged estimate of each frame, which we denote as $\hat{F_0}_{k,i}$.  
%In this paper, we use the PEFAC method \cite{} for obtaining the pitch estimation $\hat{F_0}_{ik}$. 
%The estimated speaker location, the corresponding extracted speaker identification and the associated extracted sound form a set of features of the candidate speaker, i.e. $\mathrm{x}_{k,i} \triangleq \{ \hat{\theta}_{k,i}, \hat{F_0}_{k,i}, y_{k,i}  \}$. Then we can treat the feature set as the speaker state and thereby tracking speakers via a proposed extension to the GLMB filter. 
%

From (\ref{eq:doaset}) and (\ref{eq:soundsep}), the vector of associated location, pitch and sound of each candidate speaker at frame $k$ form a multi-feature observation ${z}_{k,i} \triangleq ( \hat{\varsigma}_{k,i}, \hat{{F_0}}_{k,i}, \hat{s}_{k,i} ) $. 
%
%\begin{equation} \label{eq:multiFeatureObserv}
%{z}_{k,i} \triangleq ( \hat{\varsigma}_{k,i}, \hat{{F_0}}_{k,i}, \hat{s}_{k,i} ) . 
%\end{equation}
%
The multi-target multi-feature observation is thus 
\begin{equation} \label{eq:multiTargetMultiFeatureObserv}
Z_k \triangleq \{ z_{k,i} ~|~ i = 1,...,N_k \}  , 
\end{equation}
where $Z_k  = \emptyset$ when $N_k=0$.

Instead of using the location estimates alone, we jointly extract and track the location, pitch and sound features in the extended multi-feature GLMB filter as follows.

\section{Multi-feature GLMB} 
\label{sec:eGLMB}

The multi-feature GLMB random finite set (RFS) $\mathbf{X} \triangleq \{ (\mathrm{x}_i, \ell_i) ~|~ i \in \mathbb{N} \} $ is a labeled RFS with state space $\mathbb{X}$ (here $\mathrm{x}_i \triangleq ( \zeta_i, {F_0}_i, s_i ) \in \mathbb{X}$ is the multi-feature target state vector, where $\zeta_i,{F_0}_i,s_i$ denote the associated location and pitch feature states as well as the sound signal, respectively), and label space $\mathbb{L}, (\ell_i \in \mathbb{L}) $, where the labels are unique, i.e. $\ell_{i} \neq \ell_{i'},~ \forall i \neq i'$.  
Its probability density in the $\delta$-GLMB form is given as \cite{vo2013labeled} 
\begin{equation}
\mathbf{\pi }(\mathbf{X})=\Delta (\mathbf{X})
\!\!\!\! \sum_{(I,\xi )\in \mathcal{F}(%
\mathbb{L})\times \Xi }\omega ^{(I,\xi )}\delta _{I}(\mathcal{L(}\mathbf{X}))%
\left[ p^{(\xi )}\right] ^{\mathbf{X}} ,  \label{eq:generativeGLMB}
\end{equation}
where $\omega ^{(I,\xi )}$ is the probability of the hypothesis $(I,\xi )$, $I$ is a set of labels, $\xi$ represents a history of association map between targets and observations. $p^{(\xi )}$ is the probability distribution of a target state, $\Delta (\mathbf{X})$ is the distinct label indicator, $\delta _{I}(\mathcal{L(}\mathbf{X}))$ indicates whether the set of labels in $\mathbf{X}$ matches that of $I$.
The $\delta$-GLMB is completely characterized by the set of parameters $\{ ( \omega ^{(I,\xi )},  p^{(\xi )} ) : (I,\xi )\in \mathcal{F}(%
\mathbb{L})\times \Xi  \} $. 
(Reader are encouraged to read \cite{vo2013labeled,vo2014labeled, reuter2014labeled} and their references for detailed studies of the (G)LMB and $\delta$-GLMB RFS tracking filters.) 

The multi-feature GLMB recursion also consists of the multi-object ``update'' step based on Bayes inference and the Chapman-Kolmogorov \cite{gardiner1985handbook} ``prediction'' step based on the state transition models. 
%
%Instead of using location observations alone as in the standard GLMB implementations, we treat the location and pitch as independent transitioning features that are modeled with corresponding transition and likelihood functions, while the non-stationary sound signal extracted is treated as the non-transitioning feature and directly assigned with the new observation in the update. 

\subsection{Multi-feature GLMB Recursion: Update}
\label{sec:eGLMBupdate}

If the current RFS prediction density is a $
\delta $-GLMB of the form (\ref{eq:generativeGLMB}), using the current multi-feature observation $Z$ as defined in (\ref{eq:multiTargetMultiFeatureObserv}), the posterior density is a $\delta $-GLMB \cite{vo2014labeled}, i.e.  
\allowdisplaybreaks
\begin{equation}
\begin{aligned} 
 & \mathbf{\pi }\!(\mathbf{X}| Z )= \\ &
\Delta \!(\mathbf{X})\!\!\!\!\!\!\!\!\sum_{(I,%
\xi )\in \mathcal{F}\!(\mathbb{L})\!\times \!\Xi }\;\sum\limits_{\theta \in \Theta \!(I)}\!\!\!\!\omega^{\!(I,\xi ,\theta \!)\!}(Z) \delta
_{\!I\!}(\mathcal{L\!(}\mathbf{X})\!)\!\!\left[ p^{\!(\xi ,\theta )\!}(\cdot
| Z )\right] ^{\!\mathbf{X}} ,  \label{eq:PropBayes_strong0}
\end{aligned}%
\end{equation}
where $\Theta (I)$ denotes the subset of current association maps with
domain $I$, and standard derivations of $\omega ^{(I,\xi ,\theta )\!}(Z)$ and $p^{\!(\xi ,\theta )\!}(\mathrm{x},\ell |Z)$ are provided in \cite{vo2014labeled}. (For denotation simplicity we drop the subscript $k$ here.)

%
%\allowdisplaybreaks%
%\begin{eqnarray}
%\omega ^{(I,\xi ,\theta )\!}(Z)\!\!\! &\propto &\!\!\!\omega ^{(I,\xi
%)}[\eta _Z^{(\xi ,\theta )}]^{I}  \label{eq:PropBayes_strong1} \\
%p^{\!(\xi ,\theta )\!}(\mathrm{x},\ell |Z)\!\!\! &=&\!\!\!\frac{p^{(\xi )}(\mathrm{x},\ell
%)\psi _Z(\mathrm{x},\ell ;\theta )}{\eta _Z^{(\xi ,\theta )}(\ell )}
%\label{eq:PropBayes_strong3} \\
%\psi _Z(\mathrm{x},\ell ;\theta )&=&\!\!\!\!\! \left\{
%\begin{array}{ll}
%\!\!\!\! \frac{p_{D}(\mathrm{x},\ell ) \mathrm{g} ({z}_{\theta (\ell )}|\mathrm{x},\ell )}{\kappa ({z}_{\theta (\ell
%)})}, \text{if }\theta (\ell )>0 \\
%1-p_{D}(\mathrm{x},\ell ), \text{if }\theta (\ell )=0%
%\end{array}%
%\right.  \label{eq:PropConj5}  \\
%\eta _Z^{(\xi ,\theta )}(\ell )\!\!\! &=&\!\!\!\left\langle p^{(\xi
%)}(\cdot ,\ell ),\psi _Z(\cdot ,\ell ;\theta )\right\rangle 
%\label{eq:PropBayes_strong2}
%%\\
%%g(Z|\mathbf{X}) &=& e^{-\left\langle \kappa ,1\right\rangle }\kappa
%%^{Z}\sum_{\theta \in \Theta (\mathcal{L(}\mathbf{X}))}\left[ \psi
%%_{\!Z}(\cdot ;\theta )\right] ^{\mathbf{X}}  \label{eq:RFSmeaslikelihood0}
%\end{eqnarray}
%
Following the definitions in \cite{vo2014labeled}, clutter is assumed Poisson with an average of 0.044 clutter points per scan, i.e. the localization method in Section \ref{sec:mcc-phat} produces almost clean location estimates in low reverberation. 
The probability of a target state being detected is $p_D = 0.98 \mathcal{N}({F_0}; 280, 30^2)/\mathcal{N}(280; 280, 30^2)$.
%The standard inner product notation is defined as $\left\langle f,g\right\rangle \triangleq \int f(\mathrm{x})g(\mathrm{x})d\mathrm{x} $.
%

In this paper, $\mathrm{g}({z}_{\theta (\ell )}|\mathrm{x},\ell )$ denotes the multi-feature likelihood for the measurement ${z}_{\theta (\ell )} \in Z $ being generated by $(\mathrm{x},\ell) = ( ( \zeta, {F_0}, s ), \ell)$, where $s = \hat{s}_{\theta(\ell)}$ after update. Sound separation for respective speakers over time is achieved by concatenating sound signals $s$ of the same target label. 
Assuming that the transitioning features (location and pitch) are statistically independent, the proposed multi-feature likelihood function is:
%%
%\begin{equation}
%\mathrm{g}({z}_{\theta (\ell )}|\mathrm{x},\ell ) = \frac{\mathrm{g}_{\varsigma} + \mathrm{g}_{{F_0}}}{1+ \mathrm{g}_{\varsigma} \cdot \mathrm{g}_{{F_0}}} , 
%\end{equation} 
%%
%
\begin{equation}
\mathrm{g}({z}_{\theta (\ell )}|\mathrm{x},\ell ) \triangleq \mathrm{g}(\hat{\varsigma}_{\theta(\ell)} | \zeta, \ell ) \cdot \mathrm{g}(\hat{{F_0}}_{\theta(\ell)} | {F_0}, \ell ) , 
\end{equation} 
where $\mathrm{g}(\hat{\varsigma}_{\theta(\ell)} | \zeta, \ell ) = \mathcal{N}(\hat{\varsigma}_{\theta (\ell )} ; \zeta, \sigma_\varsigma^2)$ and $\mathrm{g}(\hat{{F_0}}_{\theta(\ell)} | {F_0}, \ell ) = \mathcal{N}(\hat{{F_0}}_{\theta (\ell )} ; {F_0}, \sigma_{F_0}^2) $ in this paper. 
%
%%
%\begin{equation}
%%\mathrm{g}_{\varsigma} = 
%\mathrm{g}(\hat{\varsigma} | \zeta, \ell ) = \mathcal{N}(\hat{\varsigma} ; \zeta, \sigma_\varsigma^2) , 
%\end{equation}
%%
%%and
%%
%\begin{equation}
%%\mathrm{g}_{{F_0}} = 
%\mathrm{g}(\hat{{F_0}} | {F_0}, \ell ) = \mathcal{N}(\hat{{F_0}} ; {F_0}, \sigma_{F_0}^2) , 
%\end{equation}
%%
$\sigma_\varsigma = 2^{\circ}$ and $\sigma_{F_0} = 10$Hz are the standard deviations of the observation of the location and pitch, respectively. 
After update, the maximum \textit{a posteriori} (MAP) estimate of the cardinality (number of speakers) is chosen, and the highest weighted corresponding hypothesis is used for the multi-target multi-feature tracking results.

\subsection{Multi-feature GLMB Recursion: Prediction}
\label{sec:GLMBprediction}

If the current RFS filtering density from its previous update step is a $%
\delta $-GLMB of the form (\ref{eq:generativeGLMB}), the prediction density to the next time is a $\delta $-GLMB given as \cite{vo2014labeled}
\begin{equation}
\begin{aligned}
\mathbf{\pi }_{\! +} & (\mathbf{X}_{\!+\!})
=   \Delta(\mathbf{X}%
_{\!+})\!\!\!\!\!\!\!\sum_{(I_{+},\xi )\in \mathcal{F}(\mathbb{L}_{+})\times
\Xi }\!\!\!\!\omega _{+}^{(I_{+},\xi )}\delta _{I_{+\!}}(\mathcal{L(}\mathbf{X}%
_{\!+}))\!\left[ p_{+}^{(\xi )\!}\right] ^{\!\mathbf{X}_{+}} , 
\label{eq:PropCKstrong1}
\end{aligned}%
\end{equation}
%
%\begin{equation}
%\mathbf{\pi }_{\!k\!+\!1|k\!}(\mathbf{X})=\Delta \!(\mathbf{X})\!\!\!\!\!\!\!\!\!\!\!\!\!\sum_{(I,\xi
%)\in \mathcal{F}(\mathbb{L}_{0:k+1})\times \Theta
%_{0:k}}\!\!\!\!\!\!\!\!\!\!\!\!\omega _{k+1|k}^{(I,\xi )}\delta _{\!I}(%
%\mathcal{L(}\mathbf{X}))\!\left[ p_{k\!+\!1|k}^{(\xi )}\right] ^{\!\mathbf{X}%
%}  \label{eq:deltaGLMBprediction}
%\end{equation}
%
%
where standard derivations of $\omega_+ ^{(I_+,\xi )}$ and $p_{+}^{(\xi )}(\mathrm{x},\ell )$ can be found in \cite{vo2014labeled}.
%
%\allowdisplaybreaks%
%\begin{eqnarray}
%\!\!\!\omega_+ ^{(I_+,\xi )}\!\! &=&\!\!\omega _{S}^{(\xi )}(I_{+}\cap
%\mathbb{L}) w_{B}(I_{+}\cap \mathbb{B})  \label{eq:PropCKstrong2} \\
%\!\!\!p_{+}^{(\xi )}(\mathrm{x},\ell )\!\! &=&\!\!1_{\mathbb{L}}(\ell )p_{S}^{(\xi
%)\!}(\mathrm{x},\ell )+1_{\mathbb{B}\!}(\ell )p_{B}(\mathrm{x},\ell )  \label{eq:PropCKstrong3}
%\\
%\!\!\!\omega _{S}^{(\xi )}(L)\!\! &=&\!\![\eta _{S}^{(\xi
%)}]^{L}\sum_{I\supseteq L}[1-\eta _{S}^{(\xi )}]^{I-L}\omega ^{(I,\xi )}
%\label{eq:PropCKstrongws} \\
%\!\!\!p_{S}^{(\xi )}(\mathrm{x},\ell )\!\! &=&\!\!\frac{\left\langle p_{S}(\cdot
%,\ell ) \mathrm{f}(\mathrm{x}|\cdot ,\ell ),p^{(\xi )}(\cdot ,\ell )\right\rangle }{\eta
%_{S}^{(\xi )}(\ell )}  \label{eq:PropCKstrong4} \\
%\!\!\!\eta _{S}^{(\xi )}(\ell )\!\! &=&\!\!\left\langle p_{S}(\cdot ,\ell
%),p^{(\xi )}(\cdot ,\ell )\right\rangle   \label{eq:PropCKstrong_eta} 
%\end{eqnarray}
%
$[\cdot]_+$ stands for prediction. The survival probability is $p_S(\cdot, \ell) = 0.75$.

Using the assumption that the transitioning features are statistically independent, the proposed state transition function for the multi-feature GLMB is:
\begin{equation}
\mathrm{f}(\mathrm{x}|\cdot ,\ell ) = 
1_\mathrm{x}(\zeta) \cdot \mathrm{f}(\mathrm{\zeta}|\cdot ,\ell ) 
~\cdot~ 1_\mathrm{x}({F_0}) \cdot \mathrm{f}({F_0}|\cdot ,\ell ) ,
\end{equation}
where the inclusion function is defined as 
\begin{equation}
1_{Y}(X)\triangleq \left\{
\begin{array}{l}
1,\text{ if } X ~\text{is included in}~ Y \\
0,\text{ otherwise} . %
\end{array}%
\right.  
\end{equation}

We assume the motion of the speaker DOA follows the Langevin process \cite{vermaak2001nonlinear, ward2003particle, ma2006tracking}, which is also a first-order Markov model:
\begin{equation}
\label{eq:langevin1}
\mathrm{f}(\mathrm{\zeta}|\zeta' ,\ell ) = 
\begin{bmatrix}
1 &  t_\Delta \\
0 & e^{-\beta_{\zeta} \cdot t_\Delta} 
\end{bmatrix} \cdot \zeta' 
+ w_{\zeta} \cdot 
\begin{bmatrix}
0 \\ 
\sigma_{\zeta} ~ \sqrt[]{1-e^{-2 \beta_{\zeta} \cdot t_\Delta}}
\end{bmatrix} , 
\end{equation}
$ \zeta = [\varsigma, \dot{\varsigma}]^{T} $, $\dot{\varsigma}$ is the velocity of DOA $\varsigma$. $t_\Delta = 0.1$s is the time step, $w_{\zeta}$ follows the normal distribution, i.e. $w_{\zeta} \sim \mathcal{N} (\cdot; 0,1)$. Model parameters $\beta_{\zeta} = 0.2 \mathrm{s}^{-1}$ and $\sigma_{\zeta} =10^{\circ}/\mathrm{s}$ are respectively the rate constant and the steady-state root-mean-square velocity for the random motions of speakers. 

We also assume that the pitch of a speaker follows a simple normal distribution around its previous estimate. Thus the state transition function for pitch is:
\begin{equation}
\mathrm{f}({F_0}| \mathrm{F_0}' ,\ell ) = \mathcal{N}({{F_0}} ; {F_0}', \tilde{\sigma}_{F_0}^2)   , 
\end{equation}
where $\tilde{\sigma}_{F_0} = 30$Hz is the standard deviation for the transition of pitch. 
Adaptive measurement-driven target births are generated \cite{reuter2014labeled, lin2016measurement}. New target births are assumed to follow normal distributions around the previous measurement, where the standard deviation is $5^{\circ}$ for the DOA, and $30$Hz for the pitch, respectively. 
The non-stationary sound signals are treated as the non-transitioning feature, thus targets carry no sound in prediction until the next update step of the  multi-feature GLMB recursion.

\section{Numerical Studies} \label{sec:GLMBmultiResults}

\subsection{Experiment Setup}

This section verifies and demonstrates the performance of the proposed multi-feature GLMB framework in the scenario of three speakers. 

\begin{figure}[h]
\centering
\includegraphics[width=8.5cm]{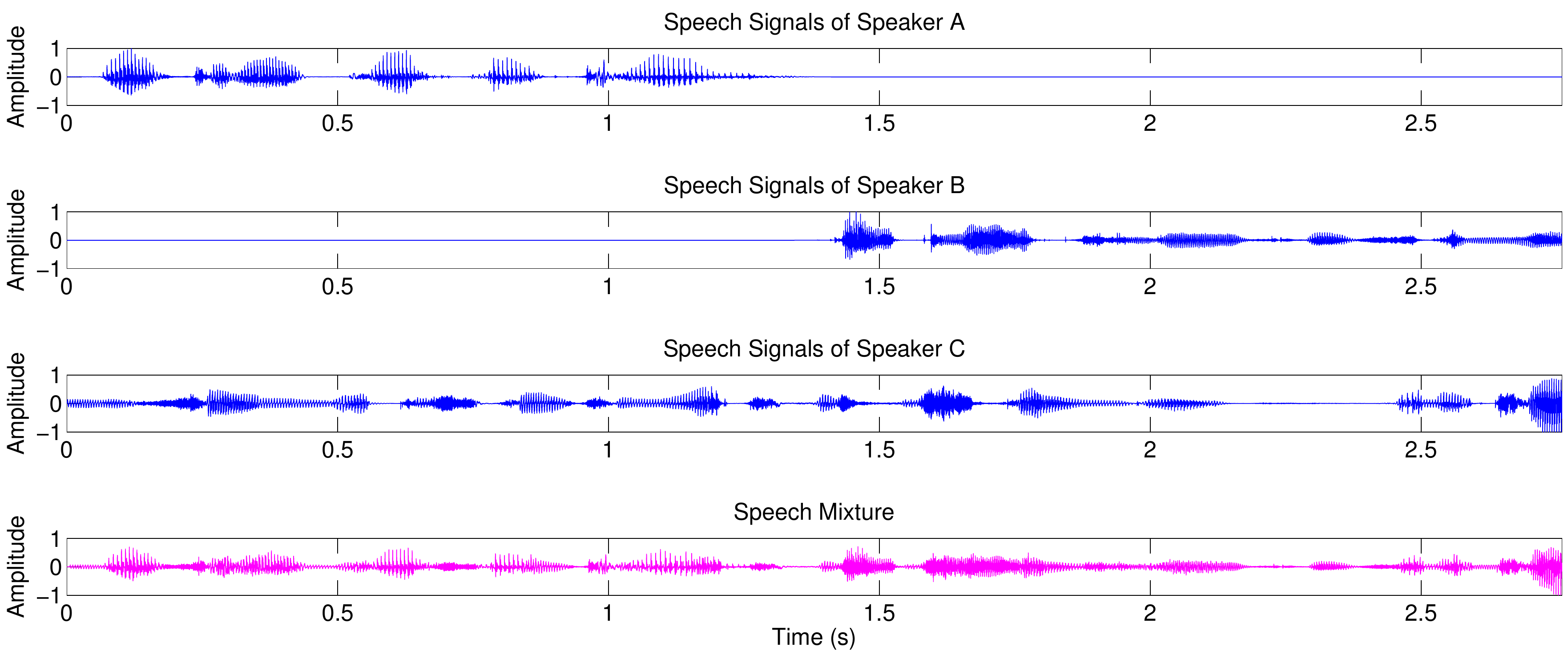}
\centering
\caption{Ground truth (top three panels) of the normalized speech signals of three speakers (one male and two female), and their mixture at one of the microphones (bottom panel). }
\label{fig:groundTruth}
\end{figure}

The setup is as shown in the left panel of Fig.~\ref{fig:GLMBmulti_Overview}, where the room dimensions are $3.4(W)\times7.6(L)\times2.7(H)\mathrm{m}^3$, the microphone array locates at [1.2, 3.9, 1.5]m, which is composed of $M=8$ microphones evenly distributed on a circle with a diameter of $0.1$m. 
For clarity, we choose an anechoic scenario that Speaker A (male) and B (female) both locate at DOA of $232.1^\circ$ %[0.5, 3, 1.5]m
while Speaker C (female) moves from DOA of $40^{\circ}$ to $75^{\circ}$, %locates at %[2.5, 6, 1.5]m. DOA of $58.2^\circ$ 
with respect to the center of the microphone array.
%Impact of reverberation is simulated using the image source method (ISM) \cite{lehmann2008prediction}.
%%
%\begin{figure}[!]
%\centering
%\includegraphics[width=0.48\textwidth]{Fig/RoomSetupGLMBmult3D.pdf}
%\centering
%\caption{Room setup, locations of microphones and speakers. }
%\label{fig:RoomSetupGLMBmult}
%\end{figure}
%%
%
Fig.~\ref{fig:groundTruth} plots the normalized ground truth speech signals of respective speakers as well as their mixture captured by one of the microphones. 
%We investigate the case where Speaker A and B  talking as the Speaker A stops talking. Speaker C resides at different locations and talks independently. 
Obviously, using location (DOA) information alone, standard implementations of tracking methods can only take Speaker A and B as a same speaker. 
(The scenario when closely located speakers talk concurrently is not in the scope of this paper.)
% We also investigate the performance of proposed method in presence of reverberation using the image source method (ISM) \cite{lehmann2008prediction}. 
%

\subsection{Test Results}
\begin{figure}[!]
\centering
\includegraphics[width=0.48\textwidth]{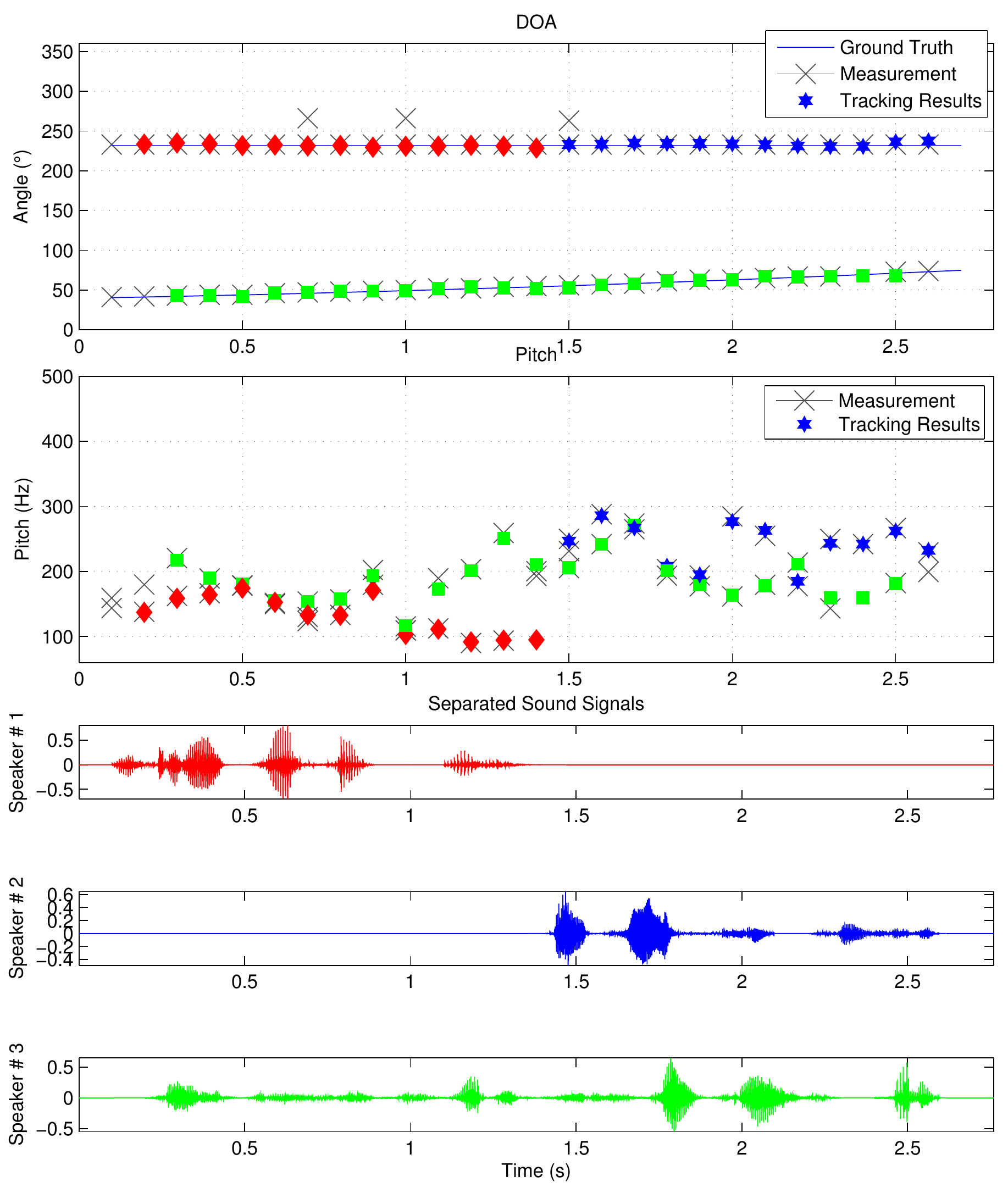}
\centering
\caption{Joint tracking and separation results from proposed methods. Top two panels show the estimation and tracking results of speakers' location and pitch. Bottom three panels show the corresponding separated sound signals. }
\label{fig:eGLMBresults}
\end{figure}

Fig.~\ref{fig:eGLMBresults} provides the ground truth locations, estimated speaker locations, pitch and separated sound signals. 
The top panel depicts the ground truth locations in straight line segments, our estimated locations in symbol ``$\times$'' and tracking results in solid colored symbols. Different colored symbols represent different speakers. From the ground truth, there are two separate lines of locations. Thus using location information alone, apparently the tracking filters can only detect two speakers. 
However, by considering also the pitch information, our proposed method has correctly found three speakers. 
The second top panel shows the pitch estimates and tracking results associated with the location estimates and tracking results in the top panel. 
We can see in these two panels that the associated location and pitch estimates have spurious errors that do not follow consistent kinematic patterns over time, thus are filtered by the GLMB tracker. 
We can also see that the tracking filter requires two time steps to confirm one new track. This is reasonable as we use the measurement-driven birth model \cite{lin2016measurement} for adaptive target births. 
The pitch estimates of different speakers  fluctuate at different levels over time, and there is a significant jump in pitch level at time of around $1.4$s, which helps the tracker to confirm a new speaker starting at $1.5$s.
The bottom three panels of Fig.~\ref{fig:eGLMBresults} plots the extracted sound signals for respective speakers. Comparing with Fig.~\ref{fig:groundTruth}, we can see that most of speech signals are recovered for each speaker. %Due to the beamforming method, there are fluctuations in amplitudes for speaker $3$ as its location changes. 
Thus our proposed multi-feature GLMB tracking-and-separation method can jointly track and separate multiple speakers.  
%
% observed speaker locations and pitches with symbol ``$\times$'' using the feature extraction methods in Section \ref{sec:features}. The implemented localization method produces clean location estimates and fluctuating pitches.
%%
%We can also see that the location estimates alone seem to indicate only two speakers at different locations. However, their pitches are different and there is an abrupt change in speaker pitch estimate at about $1.4$s when Speaker A stops talking while Speaker B starts talking. 
%%
%Therefore, using the proposed multi-feature GLMB method to jointly track location and pitch of speakers, we can obtain the tracking results as shown in Fig.~\ref{fig:eGLMBresults}. Tracks that belong to respective speakers are depicted in different colors. It correctly identifies and clearly shows three tracks, i.e. three speakers. 
%

\begin{figure}[!]
\centering
\includegraphics[width=0.48\textwidth]{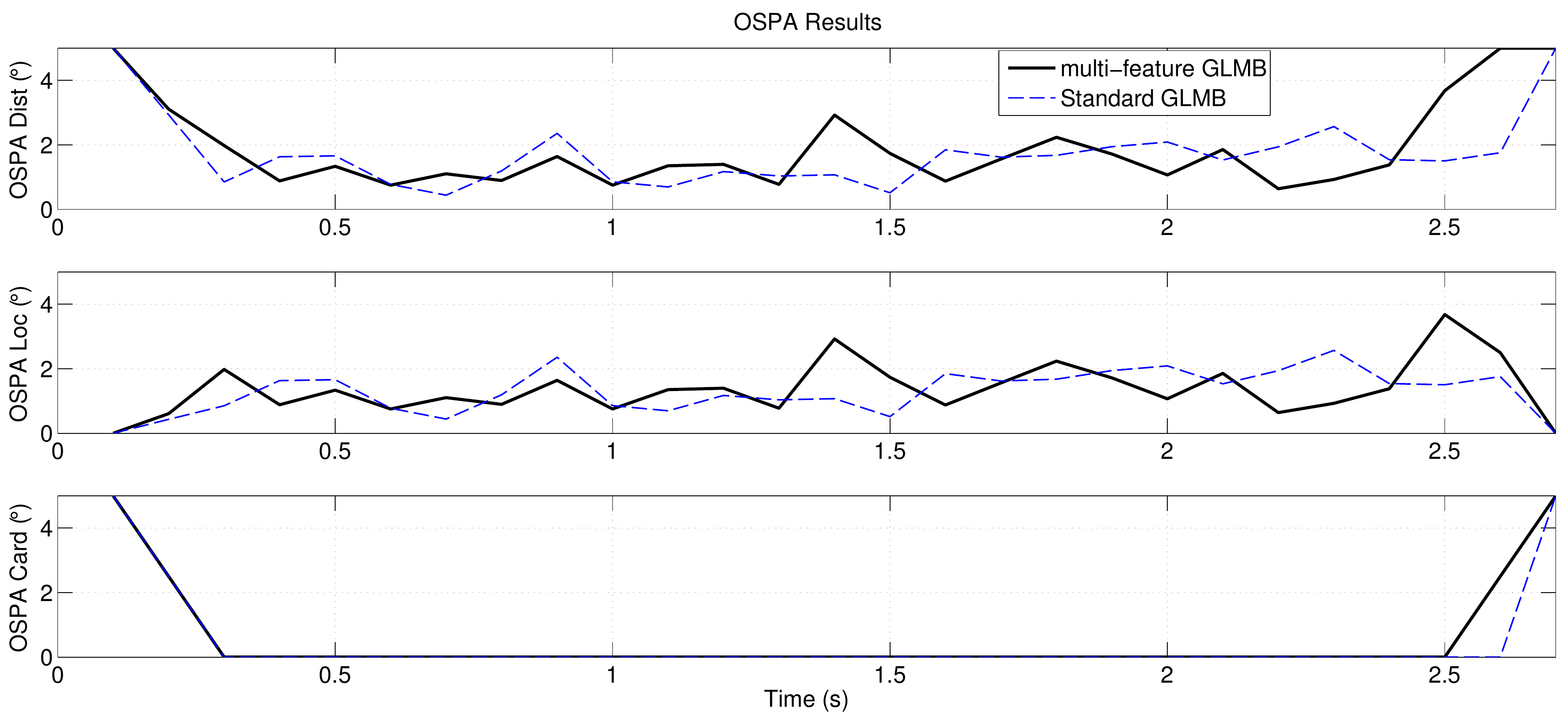}
\centering
\caption{OSPA measure of the DOA tracking results, i.e. the overall OSPA errors (top), the contribution of DOA errors (middle), and the contribution of cardinality errors (bottom).}
\label{fig:ospaDOA}
\end{figure}

The location tracking accuracy is evaluated using the Optimal Sub-pattern Assignment (OSPA) metric \cite{schuhmacher2008consistent}, with the cut-off parameter of $5^{\circ}$ and the order parameter of $1$. Thus cardinality estimation error of 1 out of 2 contributes to an OSPA error of $\frac{5}{2} ^{\circ}$. 
Fig.~\ref{fig:ospaDOA} shows that the overall OSPA location tracking errors are within $5^\circ$, 
and the multi-feature GLMB achieves comparable location tracking accuracy with the standard GLMB.
%
%
%The extracted sound signals for respective speakers are plotted in the bottom three panels of Fig.~\ref{fig:eGLMBresults}. 
% Fig.~\ref{fig:separatedSound}. 
%Comparing with Fig.~\ref{fig:groundTruth}, we can see that most of the speech signals are extracted for each speaker. 
%
%\begin{figure}[!]
%\centering
%\includegraphics[width=0.48\textwidth]{Fig/GLMB_MCCPHAT_BF_Pefac_SeparatedSpeakers.pdf}
%\centering
%\caption{Separated speech signals ($T_{60}=0.2s$). }
%\label{fig:separatedSound}
%\end{figure}
%%

The quality of the separated sound signals are evaluated using the PEASS metric \cite{emiya2011subjective}, compared with the ground truth signals. The results are provided in Tab.~\ref{tab:PEASS}. 
We also compare the performance with two blind speech separation methods, i.e. the Underdetermined Convolutive Blind Source Separation (UCBSS) \cite{reju2010underdetermined} and the Degenerative Unmixing Estimation Technique (DUET) \cite{yilmaz2004blind}. 
We can see that using the blind separation techniques, the speaker 1 and speaker 2 are regarded as one speaker. Thus the separated sound signals for speaker $<1,2>$ are compared with the mixture of Speaker A and Speaker B. 
In general the DUET and UCBSS methods obtain close Overall Perceptual Scores (OPS). The DUET method seems to provide more consistent performance than UCBSS when comparing the Target-related Perceptual Score (TPS) and the Artifacts-related Perceptual Scores (APS), but UCBSS has significantly higher Interference-related Perceptual Score (IPS) than DUET.  
Overall, our proposed method provides consistent and superior performance for the three separated speakers, according to all the perceptual scores.  

\begin{table} [!h]
\centering
\caption{PEASS evaluation results for speech separation, using the proposed method, and the UCBSS, DUET methods.}
\begin{tabular}{| c || c | c | c | c | c | c |}	
\hline
Method & Speaker  & OPS	& TPS & IPS	& APS 
\\\hline%{|=||=|=||=|=|=|=|}
\multirow{ 2}{*}{Proposed}  & $1$ & 48.75 &	57.03	& 71.19 & 49.11 \\
& $2$	& 32.69 & 29.35 & 72.06 & 35.61 \\
& $3$	& 36.02 & 35.73 & 65.65 & 37.71
\\\hline
\multirow{ 2}{*}{UCBSS}  & $<1,2>$ & 18.66 &	45.84 & 43.21 & 24.33 \\
& $3$ & 25.00 & 6.10 & 83.97 & 3.50
\\\hline
\multirow{ 2}{*}{DUET}  & $<1,2>$ & 18.73 &	38.82 & 16.38 & 50.43 \\
& $3$ & 24.97 &	51.16 & 32.40 &	44.32	
\\\hline
\end{tabular}
\label{tab:PEASS}
\end{table}
%
%PESQ metric \cite{rix2001perceptual, hu2008evaluation}
%%, and the results are provided in Table.~\ref{tab:PEASS}. 
%Fig.~\ref{fig:OPSbar} shows that the overall perceptual scores (OPS) are all over 20 for the three extracted sound signals, which indicates consistent and encouraging separation performance. 
%%
%\begin{figure}[!]
%\centering
%\includegraphics[width=0.48\textwidth]{Fig/GLMBmulti_T60leq02s.pdf}
%\centering
%\caption{OPS of speech separation results. }
%\label{fig:OPSbar}
%\end{figure}
%%

\section{Conclusion and Future Work} \label{sec:conclusion}

This paper presents the novel systematic implementation of multi-feature GLMB tracking method that not only can jointly track multiple speakers and separate sound signals from speech mixtures, but also resolve the ambiguity of location tracking when speakers locate spatially close. 
It treats the vector of candidate speaker location, pitch and sound as a multi-feature target observation and jointly extracts and tracks these features in the Bayes RFS recursion.
Experimental results demonstrate encouraging results in the studied scenario. 
For future work, further improvement is still possible, e.g. by applying more complicated microphone setup, selecting different speaker features, or improving the feature extraction methods. %(using more microphone arrays, choosing other speaker acoustic identities, and applying  superior speech separation techniques, etc.). % for sound signals and acoustic identities. 

% -------------------------------------------------------------------------
\bibliographystyle{IEEEbib}
\bibliography{Bibliography}

\end{document}